\documentclass[aps,prl,twocolumn,groupedaddress,showpacs]{revtex4}

\newcommand{\calion}[1]{$^{#1}$Ca$^+$}

\usepackage{latexsym} \usepackage{amssymb} \usepackage{amsmath}
\usepackage[dvips]{graphicx}
\usepackage[dvips]{color}

\begin{document}


\title{Observation of Three-dimensional Long-range Order\\
 in Smaller Ion
  Coulomb Crystals in an rf Trap}


\author{A.~Mortensen}
\author{E.~Nielsen}
\author{T.~Matthey}
\author{M.~Drewsen}\email[E-mail: ]{drewsen@phys.au.dk}

\affiliation{QUANTOP - Danish National Research Foundation Center for
  Quantum Optics, \\
  Department of Physics and Astronomy, University of Aarhus, DK-8000
  Aarhus C, Denmark}


\date{\today}

\begin{abstract}

  Three-dimensional long-range ordered
  structures in smaller and near-spherically symmetric Coulomb crystals of $^{40}$Ca$^{+}$ ions confined in a
  linear rf Paul trap have been observed when the number of ions
  exceeds $\sim$1000 ions. This result is unexpected from ground state
  molecular dynamics (MD) simulations, but found to be in agreement
  with MD simulations of metastable ion configurations. Previously,
  three-dimensional long-range ordered structures have only been reported in Penning traps in
systems of $\sim$50,000 ions or more.

\end{abstract}

\maketitle

A Coulomb crystal is the solid state phase of a confined ensemble of
Coulomb interacting particles with the same sign of charge, often
referred to as a one-component plasma (OCP). Coulomb crystallization
and properties of Coulomb crystals have been studied experimentally for decades
in such a variety of systems
as 2D electron gases on super-fluid helium~\cite{Grimes:1979} and in
quantum well structures~\cite{Andrei:1988}, laser cooled ions in
traps~\cite{Birkl:1992,Mitchell:1998,Itano:1998,Drewsen:1998,Hornekaer:2001,Schatz:2001,Tan:1995,Kjaergaard:2003,Bluemel:1988}
and most recently dusty plasmas~\cite{Arp:2004}.
In nature, Coulomb crystals are presently expected to exist in
exotic dense astrophysical objects~\cite{Horn:1991}.

  Theoretically, it has been found that the thermodynamic properties
  of infinite OCPs of a single species, are fully characterized by the coupling parameter~\cite{Ichimaru:1982}
\begin{equation}
  \label{eq:1}
\Gamma
= \frac{1}{4\pi\epsilon_0} \frac{Q^2}{a_\mathrm{ws}k_B T} ,
\end{equation}
where $Q$ is the charge of the particles, $a_\mathrm{ws}$ is the
Wigner Seitz radius defined from the zero temperature particle
density $n_0$ by $4\pi a_\mathrm{ws}^3/3 = 1/n_0$. Furthermore, a
liquid-solid transition to a body centered cubic (bcc) structure is expected to occur for
$\Gamma \sim$ 170~\cite{Pollock:1973,Slattery:1980}.  For finite
OCPs simulations have shown that the situation is more complex, and the properties will depend both
on the size and shape of the ion plasma, since surface effects
cannot be neglected~\cite{Rahman:1986,Hasse:1990,Schiffer:1993,Schiffer:2002,Dubin:1988}.

Ion Coulomb crystals, which for more than a decade have been realized
with laser-cooled ion plasmas confined by electromagnetic fields in
Penning traps~\cite{Tan:1995,Itano:1998,Mitchell:1998} or in
radio-frequency (rf) traps~\cite{Birkl:1992,Drewsen:1998,Hornekaer:2001,Kjaergaard:2003,Schatz:2001}, offer an excellent
opportunity to study finite size effects of OCPs under various
conditions.
The Coulomb crystal structures studied range from one-dimensional (1D)
long cylindrical
crystals~\cite{Birkl:1992,Drewsen:1998,Hornekaer:2001,Schatz:2001}
over 2D thin planar crystals~\cite{Mitchell:1998} to 3D spheroidal
crystals~\cite{Drewsen:1998,Tan:1995,Itano:1998,Mitchell:1998}.
The 3D spheroidal ion Coulomb crystals reported in
Refs.~\cite{Drewsen:1998,Hornekaer:2001} are composed of concentric
ion shells formed under the influence of the surface of the Coulomb
crystals. Simulations indicate that the ions form a near-2D
hexagonal short-range ordered structure within each
shell~\cite{Hasse:1990}. Similarly, shell and short-range order have
recently been observed in dusty plasma experiments~\cite{Arp:2004}.

Observations of three-dimensional long-range order in Coulomb
crystals have previously only been reported in the case of $\sim$50,000
or more laser-cooled ions in a Penning
trap~\cite{Tan:1995,Itano:1998,Mitchell:1998}. In contrast to
Penning traps, in rf traps Coulomb crystals undergo strong
quadrupole deformations at the frequency of the applied rf field due
to the so-called micro-motion~\cite{Bluemel:1989} of the ions.
Since this motion is known to produce
heating~\cite{Bluemel:1989,Schiffer:2001}, it has not been
obvious that three-dimensional long-range order could be obtained in
such traps.

In this Letter, we present observations of long-range structure in
Coulomb crystals of $^{40}$Ca$^{+}$ ions confined in a linear rf Paul
trap. By varying the number of ions in near-spherically symmetric
crystals, we have shown that bcc structures indeed can be observed 
in such traps even with the number of ions being below a
thousand. 

The linear Paul trap used in the experiments is described in detail
in Ref.~\cite{Drewsen:2003}, so here we give only a brief
description of its basic properties.  The linear Paul trap consists
essentially of four circular electrode rods in a quadrupole
configuration. Radial confinement of the ions is obtained by
applying a sinusoidally time-varying rf-potential $U_\mathrm{rf}\cos
\Omega_\mathrm{rf} t $ to
diagonally opposite electrodes and the same time varying potential,
but with a phase shift of $\pi$, to the two remaining electrode rods.
Static axial confinement is achieved by having all electrode rods
sectioned into three parts with the  eight end-pieces kept at a
positive dc potential $U_\mathrm{end}$ with respect to the four center
pieces.  The electrode rod radius is 2.0 mm, and the closest distance
from the trap axis to the electrodes $r_0$ is 3.5 mm. Each sectioned
rod has a 5.4 mm long center-piece and 20.0 mm long end-pieces. The
frequency of the rf field is in all experiments $\Omega_\mathrm{rf}=2\pi\times3.88$ MHz.
With the definition given in Ref.~\cite{Drewsen:2000} of the relevant
stability parameters $q$ and $a$ for the linear Paul trap under the
present operation conditions, the geometry of the trap leads to
$q=6.6\times10^{-4}U_\mathrm{rf}[V]$ and $a=5.5\times10^{-4}U_\mathrm{end}[V]$. In the
experiments the rf-amplitude $U_\mathrm{rf}[V]$ is maximally 500 V,
corresponding to $q_\mathrm{max}=0.33$. This value is small enough that in
all the experiments a harmonic pseudopotential with rotational
symmetry with respect to the trap axis and a uniform zero
temperature ion density given by
$n_\mathrm{theo}=\epsilon_0 U_\mathrm{rf}^2/m
r_0^4\Omega_\mathrm{rf}^2=1.47\times10^3U_\mathrm{rf} [V]^2$
cm$^{-3}$ can be assumed ($\epsilon_0$ is the vacuum permittivity and
$m$ the mass of the ion).

The \calion{40} ions are produced isotope-selectively by
resonance-enhanced photo-ionization of atoms from an effusive beam of
naturally abundant calcium~\cite{Kjaergaard:2000,Mortensen:2004}.
Doppler laser cooling is achieved along the trap center axis by using
counter propagating laser beams tuned to the $4 S_{1/2}$--$4P_{1/2}$
transition at 397 nm and with a repumper beam on the
$3D_{3/2}$--$4P_{1/2}$ transition at 866 nm. By this choice of cooling
geometry the fluorescence level of the individual ion will not
significantly be effected by the rf-induced micro motion (see Fig.~1),
as this motion is perpendicular to the direction of the cooling laser
to better than 1 mrad. The radial motion of the 
ions is cooled indirectly through the Coulomb coupling of the radial
and axial degrees of freedom.

Images of the positions of the ions in
the trap are obtained by detecting the 397 nm light spontaneously
emitted during the laser cooling process by a CCD-camera equipped
with an image intensifier and a 14$\times$ magnification lens
system. As indicated in Fig.~1(a), the imaging optic is situated
such that the real crystal structure is projected to a plane
including the trap axis. Due to the rotational symmetry of the Coulomb crystal boundary around
the trap axis, we can deduce from the images  their real 3D sizes and
hence the number of ions by using the expression for the ion density
given above.

\begin{figure}[htbp]
 \centering
   \includegraphics[width=8cm,clip=true]{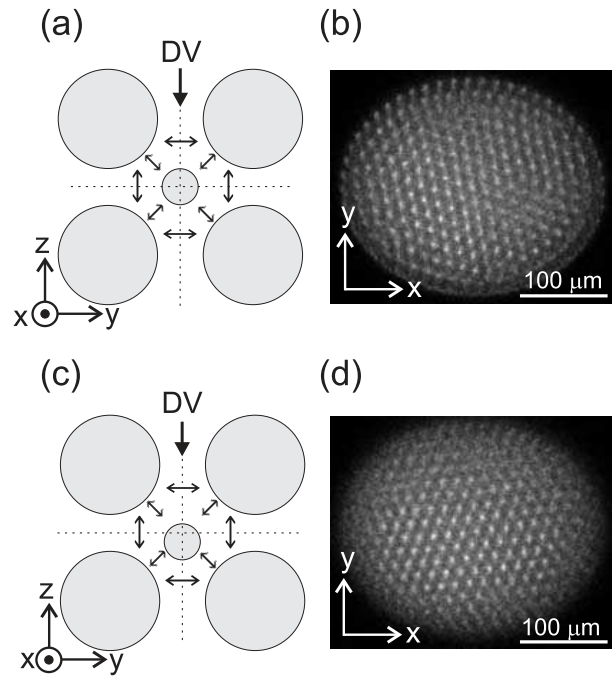}
  \caption{Observation of hexagonal structures in images of a Coulomb
crystal consisting of $\sim$2300 ions. (a) The four main circles
indicate the four cylindrical trap electrode rods viewed along the trap
axis. The small circle in the center represents a radially centered
Coulomb crystal in the trap. The small double arrows indicate the
positional dependent direction of the micro-motion of the ions in
the trap. The single arrow denotes the direction of view (DV) for
the camera system. The horizontal dashed line indicates the plane
including the trap axis in which the direction of micro-motion is
along DV. (b) Image with the Coulomb crystal radially
centered in the trap. (c) Graphical presentation of a crystal
vertically displaced. (d) Image of a vertically displaced
crystal. In both experiments: $U_\mathrm{rf}$=400 V and $U_\mathrm{end}$=15 V.
 }
  \label{fig:figure1}
\end{figure}

With an exposure time of 100 ms used in the experiments, the
quadrupole deformations of the Coulomb crystal induced by the rf field
are averaged out in the images. The direction of
the micro-motion of the individual ions is position dependent as
indicated in Fig.~1(a). As a result, only ions close to the
horizontal trap plane defined by the horizontal dashed line in Fig.~1(a) and
the trap axis, are imaged without micro-motion blurring due to
their micro-motion only being in the direction of view (DV) of the
camera system.

In Fig.~1(b), an image of a slightly prolate ion Coulomb crystal
consisting of about 2300 ions is presented. Clearly visible is a
hexagonal structure. Since the depth of focus ($\sim$50 $\mu$m) is
several times the inter-ion distance ($\sim$15
$\mu$m), the observed structure cannot originate from a single layer of ions. In order
to reveal if the ordered structure persists not only very close to
the image plane, in another experiment, the Coulomb crystal was moved down a few inter-ion distances ($\sim$40 $\mu$m) by adding a positive dc voltage to the upper two
electrode rods as sketched in Fig.~1(c). By doing this one achieves
that it is a different part of the crystal that is close to the
horizontal trap plane. As seen in the image presented in Fig.~1(d), in this
situation we still observe an ordered structure apart from close to the
edges where the fluorescence light originates from ions not being close
to the image plane. This experiment substantiates that the ordered
pattern observed indeed originates from a three-dimensional long-range
ordered structure. Furthermore from a density point of view the observed structure can not just be two-dimensional.

From the pre-knowledge that the ground state of larger ion systems
is a bcc configuration~\cite{Pollock:1973,Slattery:1980}, it is
tempting to assume that the observed structure is indeed a bcc
structure observed along the [111] direction. However, also simple
cubic (sc) and face centered cubic (fcc) structures will lead to
similar hexagonal patterns for the same direction of observation.
For a specific side length $d$ of the triangles making up the
observed structure, the three cubic structures correspond, however,
to different ion densities related by $n_\mathrm{bcc}=2n_\mathrm{sc}=4n_\mathrm{fcc}=1.089/d^3$. The
perfect agreement between the expected density of $n_\mathrm{theo}= 2.3 \pm
0.2 \times10^8$ cm$^{-3}$ with the one obtained from the images
$n_\mathrm{bcc}=2.1 \pm0.3\times10^8$ cm$^{-3}$ assuming a bcc structure,
strongly supports that the observed structure is indeed a bcc and
certainly not a sc or fcc structure. As seen from Fig.~2(a)--(c), the outer shape of the crystal is not
critical for observing the long-range structures, at least not as long
as the crystal is not too oblate or prolate.

\begin{figure}[htbp]
 \centering
   \includegraphics[width=8cm,clip=true]{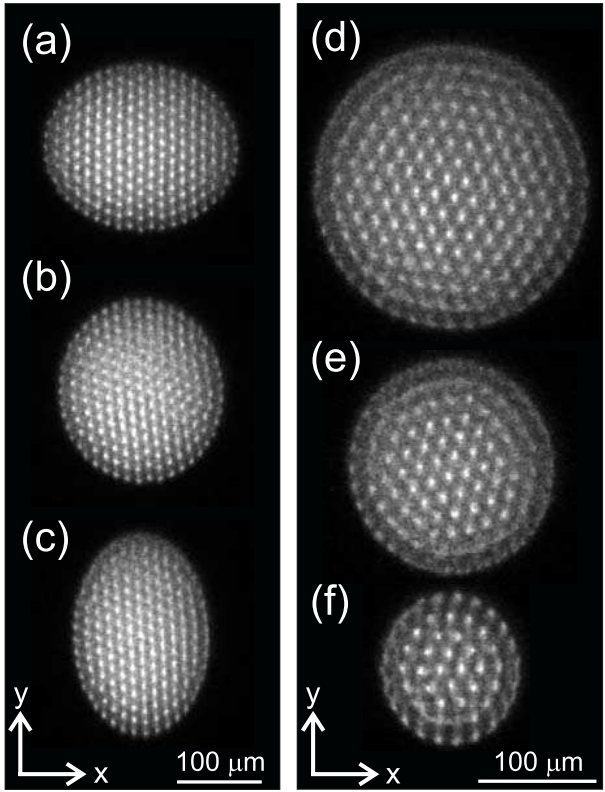}
  \caption{Images of Coulomb crystals. The first column shows
three crystal images of the same ensemble of $\sim$2000 ion with the
trap potentials being $U_\mathrm{rf}=500$ V and (a) $U_\mathrm{end} =27$ V, (b)
$U_\mathrm{end}=33$ V, and (c) $U_\mathrm{end}=40$ V, respectively. The second column
represents images near-spherical crystals for the same trap potentials
$U_\mathrm{rf}=400$ V and $U_\mathrm{end}=21$ V. The number of ions in the crystals
is (d) 1700, (e) 770, and (f) 290.
 }
  \label{fig:figure2}
\end{figure}

A series of measurements on near spheric crystals with different numbers of ions were done to
investigate the size effect of the observation of three-dimensional
long-range structures. A few resulting images of this study are
presented in Fig.~2(d)--(f). Here one observes long-range-order for as few as 770 ions (Fig.~2(e)), and even some structure is present in the core for the case of 290 ions (Fig.~2(f)).

When comparing the images in Fig.~2, it is evident that the relative
number of ions in the regular structure decreases with the size of
the crystals, as may be expected since the outer layer of the crystals
is always spheroidal shaped. Another size dependent
quantity noticed is the frequency at which ordered structures are indeed
observed: The smaller the crystals the more unlikely it is to
observe the regular structures. The rate at which the long-range
ordered structures are observed in crystals of a few thousand ions is
$\sim$0.1 Hz and the lifetime of the structure is a few hundred ms.  
In all cases, since the exposure
time of 100 ms is many orders of magnitude larger that both the
rf-period and the timescale of crystal vibrations
($\sim$$1/\omega_\mathrm{plasma}$, with $\omega_\mathrm{plasma}\sim$ $1$ MHz being the
plasma frequency), the crystal structures are at least to be
considered as metastable states.

Molecular dynamics (MD) simulations of ground state configurations
of Coulomb crystals in static spherical harmonic potentials have proven
only to have a pronounced bcc structure when the number of ions
exceeds $\sim$5000~\cite{Totsuji:2002,Hasse:2003}. However, in
contrast to such ground state simulations, a certain
thermal energy is always present in experiments, and configurations different from
the ground state may occur. In order to understand the observed
long-range-ordered structures, we have made several MD simulations using a 
static spherical harmonic potential
where we initially kept a core of the ions fixed in a bcc-structure
and cooled the surrounding ions slowly until $\Gamma >$ 100.000 was reached. 
For an ion system of $\sim$1000 ions, the excess
potential energy (relative to the ground state configuration) of such artificially created cold configurations is even with more than one tenths of the ions kept fixed in a
bcc-structure, several times smaller than the thermal energy of the
system at the typical experimental temperature of a few mK.
Simulations where all the ions were subsequently slowly heated to
$\Gamma\sim 400$ (a few
mK) furthermore showed that the artificially made configurations could
be metastable on the time scale of the exposure time of the images, $\sim$10
ms.  Figure 3(a) is an example of a constructed image
based on the results of a MD simulation of a crystal of 2685 ions
heated to $\Gamma = 400$ (temperature: 5 mK),
where initially 10\% of the ions were kept fixed in a
bcc-structure. With the 10 ms integration time of this
simulation, the central bcc-structure is seen to be well preserved, and the
image resembles very much the experimentally obtained image of a
Coulomb crystal with about the same number of ions (Fig.~3(b)). 


\begin{figure}[htbp]
 \centering
   \includegraphics[width=8cm,clip=true]{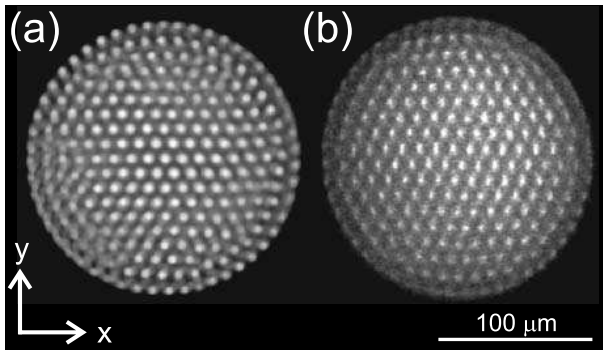}
  \caption{Images of Coulomb crystals. (a) Time averaged image
   based on data from MD simulations of Coulomb clusters with 2685
   ions at $\Gamma\sim 400$ (temperature: $\sim$5 mK). The averaging
   time is 10 ms. (b) Image from experiments with
   clusters containing  $\sim$2700 ions.
 }
  \label{fig:figure3}
\end{figure}

\begin{figure}[htbp]
 \centering
  \includegraphics[width=8cm,clip=true]{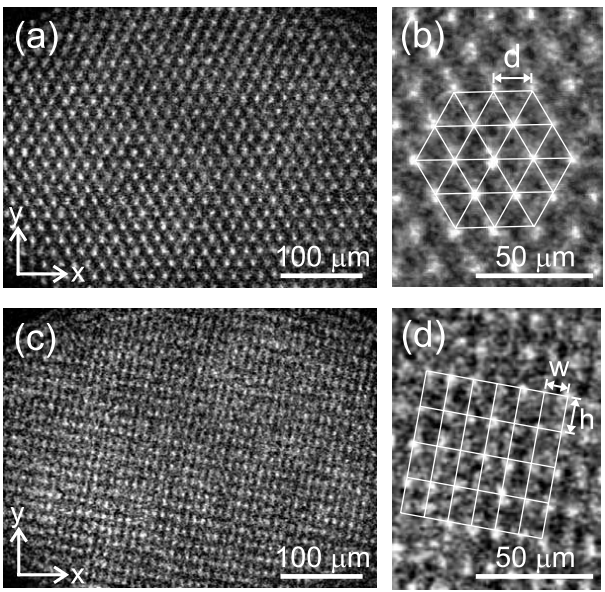}
  \caption{
Images of a cold ensemble of $\sim$13,000 ions.
(a)Visible hexagonal structures indicating a three-dimensional bcc
structure. (b) Magnification of a section of  (a). The side
length $d$ is found to be 17.2 $\mu$m. (c) Visible rectangular
structure likely indicating a slightly distorted fcc structure.
(d) Magnification of a section of (c). Here, $h = 15.2 \pm 0.5$
$\mu$m and $w= 9.71 \pm 0.2$ $\mu$m, respectively.
 }
  \label{fig:figure4}
\end{figure}

For crystals with more than
$\sim$2000 ions, images are
observed, which suggests three-dimensional long-range
ordering different from bcc. In Fig.~4, two images of the same ion ensemble
($\sim$13,000 ions) at two different instants are presented. While
the hexagonal structure of Fig.~4(a) and (b) with $d= 17.2 \pm 0.8$ $\mu$m again
is compatible with a bcc structure, the rectangular structure
observed in Fig.~4(c) and (d) cannot be interpreted as another projection
of a bcc structure, but has to relate to another structure. The
sides of the rectangle are $h = 15.2 \pm 0.5$ $\mu$m and $w= 9.7
\pm 0.2$ $\mu$m, which yields a ratio of the sides of $h/w = 1.55
\pm 0.06$. We do not find that any projection of cubic crystal
structures exactly complies with this ratio, but the fcc structure
projected in the [211] direction comes close as this has the ratio
$h/w = \sqrt{8/3} \simeq 1.63$. For the expected ion density of
$n_\mathrm{theo}= (2.2\pm 0.2)\times 10^8$ cm$^{-3}$ the corresponding side lengths
of the projected rectangles should for a fcc structure have been $h
= 15.2 \pm 0.5$ $\mu$m and $w = 9.3 \pm 0.3 $ $\mu$m. The origin of
the small discrepancy between the observed and the predicted ratio
is at present unknown, but an explanation could probably be found in a
micro-motion induced distortion of the fcc lattice. The observation
of fcc structures is not so surprising since  such structure have
already been observed for larger
near-spherical Coulomb crystals in Penning traps~\cite{Itano:1998}. Furthermore, MD simulations have as well
predicted very small differences in the potential energies of bcc
and fcc structures for larger crystals~\cite{Hasse:2003}.

The reason why the observed bcc or fcc structures always seem to
observed from a specific direction is still not completely
understood. However, the non-perfect cylindrical symmetry of the
quadrupole trap configuration and the corresponding micro-motion as
well as small patch potentials may play a role.

In conclusion, we have proven that it is possible to obtain three-dimensional 
long-range ordering in smaller ion Coulomb crystals formed in rf traps. The long-range structures are found to be consistent with metastable structures observed in MD simulations.

We acknowledge financial support from the Carlsberg Foundation as well
as from the Danish National Research Foundation: Center for Quantum
Optics QUANTOP.



\end{document}